\documentclass{jetpleng}
\usepackage{mathrsfs} 
\usepackage{cite}

\twocolumn
\lat

\newcommand{\eps}{\varepsilon}

\newcommand{\su}{\uparrow}
\newcommand{\sd}{\downarrow}

\newcommand{\ii}{{\mathrm{i}}}
\newcommand{\vk}{{\mathbf{k}}}
\newcommand{\q}{{\mathbf{q}}}
\newcommand{\p}{{\mathbf{p}}}

\usepackage{color}
\usepackage{graphicx}
\usepackage[pdfborder={0 0 0}, colorlinks=true, linkcolor=blue, citecolor=blue, urlcolor=blue]{hyperref}

\title{Superconducting order parameter structure in the nematic phase of iron-based materials}

\rtitle{Order parameter structure in the nematic phase of iron-based materials}

\sodtitle{Superconducting order parameter structure in the nematic phase of iron-based materials}

\author{M.\,M.\,Korshunov\/$^{+}$\/\thanks{mkor@iph.krasn.ru}, Yu.\,N.\,Togushova\/$^{*}$}

\rauthor{M.\,M.\,Korshunov, Yu.\,N.\,Togushova}

\sodauthor{Korshunov, Togushova}

\address{~\\$^{+}$Kirensky Institute of Physics, Federal Research Center
``Krasnoyarsk Science Center of the Siberian Branch of the Russian Academy of Sciences'', 660036 Krasnoyarsk, Russia, 660036 Krasnoyarsk, Russia\\~\\
$^{*}$ Siberian Federal University, 660041 Krasnoyarsk, Russia}

\dates{\today}{*}

\abstract{We consider the effect of the nematic order on the formation of the superconducting state in iron pnictides and chalcogenides. Nematic order with the $B_{2g}$ symmetry is modelled as the $d$-type Pomeranchuk instability and treated within the mean-field approach. Calculated nematic order parameter depends on the nematic interaction coefficient and abruptly changes with the coefficient's increase. The superconducting solution is obtained within the spin-fluctuation pairing theory. We show that the leading solution in the nematic phase has a $s_{\pi\pm}$ structure. From the critical temperature $T_c$ estimations, we conclude that the nematic superconducting state of the $s_{\pi\pm}$ type is more favorable than the usual $s_\pm$ and $d_{x^2-y^2}$ type states appearing in the absence of the nematicity.}


\begin{document}

\maketitle

\textbf{1. Introduction.}
Complicated systems are often possess several concurrent or coexisting long range orders of different nature. Iron pnictides and chalcogenides, quasi-two-dimensional systems, are an example. Multiorbital effects there lead to the appearance of an unconventional superconductivity with the order parameter structure of the $s_\pm$ type that has the opposite signs at different Fermi surface sheets, still belonging to the $A_{1g}$ representation and being the extended $s$-wave symmetry~\cite{SadovskiiReview2008,IzyumovReview2008,HirschfeldKorshunov2011,Korshunov2014eng,Sadovskii2016}. Vast amount of experimental data on the superconducting state may be explained within the framework of the spin-fluctuation mechanism of Cooper pairing that, as one of the solutions, has the $s_\pm$ state~\cite{MaitiKorshunovPRL2011}. The presence of this state is confirmed by the data on the spin-resonance peak~\cite{KorshunovEreminResonance2008,Maier2008,Korshunov2018} observed in the inelastic neutron scattering~\cite{LumsdenReview,Dai2015,Inosov2016} and by the observation of the spin exciton characteristic for the $s_\pm$ state in the Andreev reflection spectra~\cite{KorshunovKuzmichev2022}.

Experimentally discovered disparity in the resistance along the orthogonal $a$ and $b$ directions in the iron plane in the tetragonal phase of the iron pnictides~\cite{Chu2010} led to the conclusion that the $C_4$ symmetry is broken down to $C_2$ and the nematic order is formed~\cite{Fernandes2012,Fernandes2014}. The word `nematic' here is used to emphasize that the transition takes place in the electronic subsystem in contrast to the usual structural phase transition with ions moving to the new equilibrium positions. Close analog is the formation, in the disordered system of spins, of the Ising nematic order with the broken $Z_2$ symmetry. It is the difference from the usual transition to the magnetic state appearing due to the $O(3)$ symmetry breaking~\cite{Fernandes2012}. In other words, nematic phase is characterized by the non-equality in the $a$ and $b$ directions that leads to the inequality of the magnetic response, i.e. the spin susceptibility, in the orthogonal directions in the momentum space, $q_x$ and $q_y$.

Since with the lowering temperature nematic transition precedes the magnetic or superconducting transitions~\cite{Li2017,Wang2021}, we study the superconductivity on the background of the already formed nematic order. Here we analyse the role of the Fermi surface symmetry lowering from $C_4$ to $C_2$ on the superconducting gap solutions within the spin-fluctuation pairing theory~\cite{Korshunov2014eng}. Resulting solutions have the $C_2$ symmetry in agreement with the experimentally observed lowering of the gap symmetry~\cite{Sprau2017}.

\textbf{2. Model.}
We start with the Hamiltonian of the five-orbital model for iron pnictides $H_{5-orb}$~\cite{Kuroki2008,Graser2009}:
\begin{equation}
 H_{5-orb} = \sum\limits_{\vk,\sigma,l,l'} \eps_{\vk}^{l l'} d_{\vk l \sigma}^\dag d_{\vk l' \sigma},
 \label{eq:H5orb}
\end{equation}
where $d_{\vk l \sigma}^\dag$ ($d_{\vk l \sigma}$) is the creation (annihilation) operator for the electron with the momentum $\vk$, spin $\sigma$, and orbital index $l$, $\eps_{\vk}^{l l'}$ is the matrix of the single-electron energies with the chemical potential being subtracted (diagonal terms) and hopping integrals (off-diagonal elements), which values are presented in Ref.~\cite{Graser2009}.

To describe the nematic state, we follow the mean-field approach from Refs.~\cite{Yamase2005,Choudhury2022}. Two-particle nematic interaction gives the following contribution to the Hamiltonian
\begin{equation}
 H_{nem} = -\frac{1}{4}\sum\limits_{\vk,\vk',\sigma,\sigma',l,l'} V_{l l'}^{nem} f_{\vk l} f_{\vk' l'} n_{\vk l \sigma} n_{\vk' l' \sigma'},
 \label{eq:Hnem}
\end{equation}
where $n_{\vk l \sigma} = d_{\vk l \sigma}^\dag d_{\vk l \sigma}$ is the number of particle operator, $V_{l l'}^{nem}$ are the matrix elements of the interaction, $f_{\vk l}$ is the form factor. Following Refs.~\cite{Yamase2005,Choudhury2022} to model the nematic order with the $B_{2g}$ symmetry as the $d$-type Pomeranchuk instability, we set the form factor to be $f_{\vk l} = \cos{k_x} - \cos{k_y}$.

To proceed with the mean-filed theory, we write the expression $n_{\vk l \sigma} = \left< {n}_{\vk l \sigma}\right> + \delta n_{\vk l \sigma}$ with $\left< {n}_{\vk l \sigma}\right>$ being the average of the occupation number and the deviation from the average $\delta n_{\vk l \sigma}$ considered to be small. Inserting the expression into the Hamiltonian $H_{nem}$, discarding the second order terms by the deviation from the average, and omitting a constant energy shift that would be absorbed into the chemical potential, we derive
\begin{equation}
 H_{nem}^{MF} = \sum\limits_{\vk,\sigma,l} \Phi_{l} f_{\vk l} n_{\vk l \sigma}.
 \label{eq:HnemMF}
\end{equation}
Here we introduced the nematic phase order parameter
\begin{equation}
 \Phi_{l} = -\frac{1}{2}\sum\limits_{\vk',\sigma',l'} V_{l l'}^{nem} f_{\vk' l'} \left< {n}_{\vk' l' \sigma'} \right>.
 \label{eq:Phi}
\end{equation}
Note that due to the formulation of the two-particle nematic Hamiltonian  $H_{nem}$ as the density-density interaction, the mean-field Hamiltonian $H_{nem}^{MF}$ does not contain the interorbital hoppings and describes the changes of the particle density at an orbital $l$.

\begin{figure}
\centering
\includegraphics[width=0.8\linewidth]{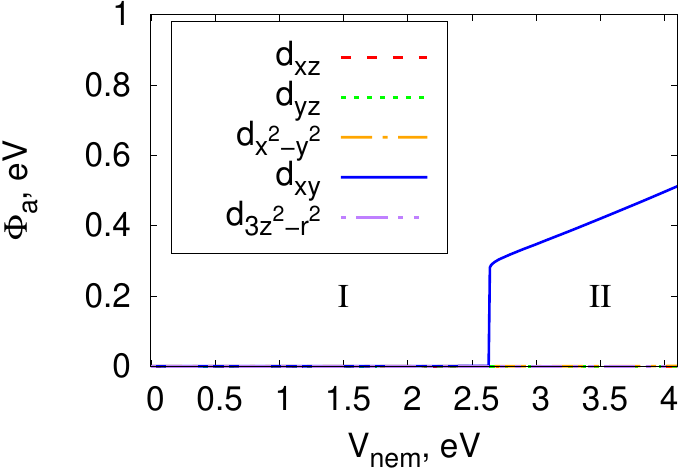}
\caption{Fig.~\ref{fig_nemPhi}. Dependence of the nematic order parameter $\Phi_a$ for the orbital $a$ on the interaction coefficient $V_{nem}$. Roman letters mark the regions where the distinct order parameter behavior is present. \label{fig_nemPhi}}
\end{figure}

\textbf{3. Nematic order parameter.}
As the first step, we find the solution for the nematic order parameter $\Phi_{l}$. To do this, we self-consistently calculate both $\Phi_{l}$ from Eq.~(\ref{eq:Phi}) and the average $\left< {n}_{\vk' l' \sigma'} \right>$. We set the matrix $V_{l l'}^{nem}$ to be equal to $\delta_{l l'} V_{nem}$ with the interaction coefficient $V_{nem}$. Since it is unknown, we treat it as a parameter. The calculated dependence of $\Phi_{l}$ for different $d$-orbitals on the coefficient $V_{nem}$ is shown in Fig.~\ref{fig_nemPhi}. Clearly, for small values of $V_{nem}$ nematic state is absent (region I). Order parameter becomes finite once $V_{nem}$ becomes larger than some value, besides, only for one of the orbitals, $d_{xy}$ (region II). Since the value of 4~eV for the interaction coefficient is quite large even compared to the Hubbard repulsion, further we restrict our consideration to regions I and II.

\begin{figure}
\centering
\includegraphics[width=0.49\linewidth]{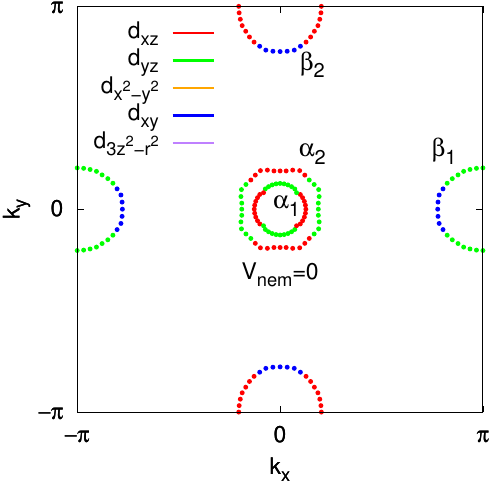}
\includegraphics[width=0.49\linewidth]{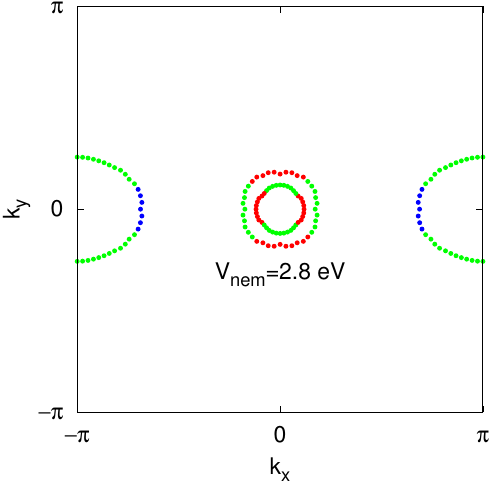}
\includegraphics[width=0.9\linewidth]{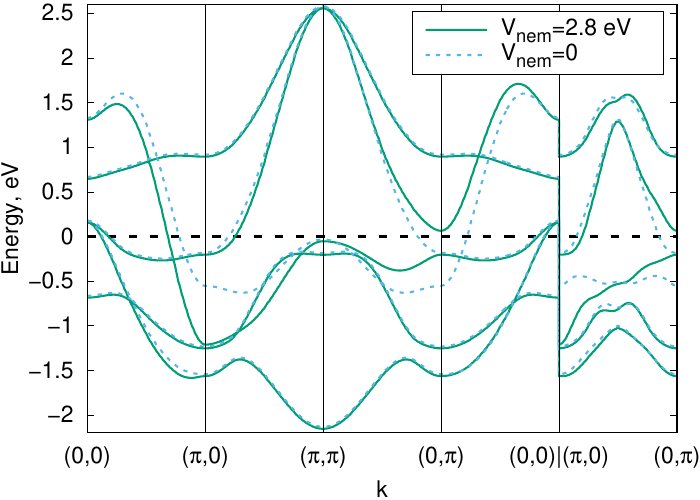}
\caption{Fig.~\ref{fig_EkFS}. Fermi surface (top) and energy dispersion along the main directions of the Brillouin zone (bottom) for the two values of the interaction coefficient $V_{nem}$. Energy is shown relative to the chemical potential. $\alpha_{1,2}$ and $\beta_{1,2}$ label the Fermi surface sheets, different colors mark the areas with the maximal contribution from the corresponding orbital. \label{fig_EkFS}}
\end{figure}

Fermi surface and energy dispersion in two different regions are shown in Fig.~\ref{fig_EkFS}. In the region II ($V_{nem} = 2.8$~eV), the breaking of the $C_4$ symmetry held in the region I ($V_{nem}=0$) is obvious. This is clearly seen in the dispersion along the $(0,\pi)-(\pi,0)$ direction.

Real part of the dynamical spin susceptibility at zero frequency $Re\chi(\q,\omega=0)$ is the central object of the spin-fluctuation pairing theory~\cite{Korshunov2014eng}. Spin susceptibility is calculated as the spin-spin correlation function
\begin{equation}
 \chi^{l l' m m'}(\q,\Omega) = \int\limits_{0}^{\beta} d\tau e^{\ii\Omega\tau} \left< T_{\tau} S_{ll'}^{+}(\q,\tau) S_{ll'}^{-}(\q,0) \right>,
 \label{eq:chi}
\end{equation}
where $\Omega$ is the Matsubara frequency, $\beta=1/T$ is the inverse temperature, $T_{\tau}$ is the time ordering operator with respect to the Matsubara time $\tau$, $S^{+}$ and $S^{-}$ are the spin operators. Averaging is done over the interacting ensemble. In the zeroth order we have
\begin{eqnarray}
 \chi^{l l' m m'}_{(0)}(\q,\Omega) &=& -T \sum\limits_{\p,\omega_n, \mu,\nu} \left[ \varphi^{\mu}_{\p m} {\varphi^*}^{\mu}_{\p l} G_{\mu \su}(\p,\omega_n) \right. \nonumber\\
 &\times& \left. G_{\nu \sd}(\p+\q,\Omega+\omega_n) \varphi^{\nu}_{\p+\q l'} {\varphi^*}^{\nu}_{\p+\q m'} \right].
 \label{eq:chi0}
\end{eqnarray}
Here $\omega_n$ is the Matsubara frequency, $\mu$ and $\nu$ are the band indices, $\varphi^{\mu}_{\vk m}$ are the coefficients of the band to orbital transformation, that is $d_{\vk m \sigma} = \sum\limits_{\mu} \varphi^{\mu}_{\vk m} b_{\vk \mu \sigma}$, where $b_{\vk \mu \sigma}$ is the electron annihilation operator in the band basis where the Green's function is diagonal, $G_{\mu \sigma}(\vk,\omega_n) = 1 / \left( \ii\omega_n - \eps_{\vk\mu\sigma} \right)$.

First we calculate $\chi^{l l' m m'}_{(0)}(\q,\Omega)$ with the Hamiltonian $H_0 = H_{5-orb} + H_{nem}^{MF}$ and than we get $\chi^{l l' m m'}(\q,\Omega)$ within the random phase approximation (RPA). Ladder approximation, RPA, is constructed with the onsite Coulomb interaction, namely, intraorbital Hubbard $U$, interorbital $U'$, Hund's exchange $J$, and the pair hopping $J'$~\cite{Castallani1978,Oles1983}.
Hamiltonian $H_{int}$ has the following form
\begin{eqnarray}
H_{int} &=& U \sum_{f, m} n_{f m \su} n_{f m \sd} + U' \sum_{f, m < l} n_{f l} n_{f m} \nonumber\\
  && + J \sum_{f, m < l} \sum_{\sigma,\sigma'} d_{f l \sigma}^\dag d_{f m \sigma'}^\dag d_{f l \sigma'} d_{f m \sigma} \nonumber\\
  && + J' \sum_{f, m \neq l} d_{f l \su}^\dag d_{f l \sd}^\dag d_{f m \sd} d_{f m \su},
\label{eq:Hint}
\end{eqnarray}
where $n_{f m} = n_{f m \su} + n_{f m \sd}$, $n_{f m \sigma} = d_{f m \sigma}^\dag d_{f m \sigma}$ is the number of particles operator on the lattice site $f$.

Sum of the ladder diagrams that include the electron-hole bubble in the matrix form $\hat\chi_{(0)}(\q,\Omega)$ gives the following expression for the spin susceptibility matrix in the RPA~\cite{Korshunov2014eng}
\begin{equation}
 \hat\chi(\q,\Omega) = \left[\hat{I} - \hat{U}_s \hat\chi_{(0)}(\q,\Omega)\right]^{-1} \hat\chi_{(0)}(\q,\Omega),
\label{eq:chiRPA}
\end{equation}
where $\hat{I}$ and $\hat{U}_s$ are the unity matrix and the interaction matrix in the orbital basis given in Ref.~\cite{Graser2009}. Later we present results for the physical susceptibility $\chi(\q,\Omega) = \sum\limits_{l,m} \chi^{l l m m}(\q,\Omega)$ analytically continued to the real frequency axis $\omega$ ($\ii\Omega \to \omega + \ii\delta$, $\delta \to 0+$).

\begin{figure}
\centering
\includegraphics[width=0.9\linewidth]{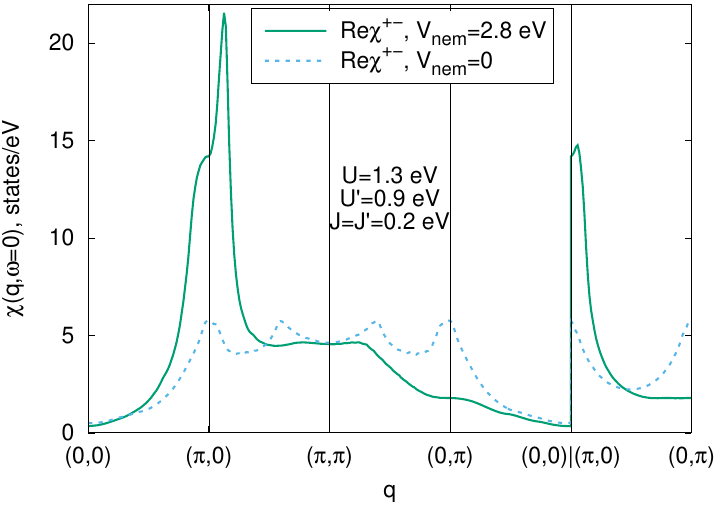}
\caption{Fig.~\ref{fig_chi}. Dependence of the real part of the spin susceptibility at zero frequency on the wave vector $\q$ calculated in RPA for the two values of the interaction coefficient $V_{nem}$. \label{fig_chi}}
\end{figure}

Breaking of the Fermi surface symmetry leads to the lowering of the symmetry to $C_2$ in the dependence of the spin susceptibility on the wave vector $\q$. This is shown in Fig.~\ref{fig_chi}. Increase of $V_{nem}$ and the following increase of the order parameter $\Phi_{l}$ leads to the rise of the peak near $\q=(\pi,0)$ in comparison with $\q=(0,\pi)$. Of course, it is related to the disappearance of the Fermi surface sheets $\beta_2$ near $(0,\pm\pi)$, in contrast to $\beta_1$ sheets near $(\pm\pi,0)$. This leads to the dominance of the scattering between the bands forming $\alpha_{1,2}$ sheets and bands forming $\beta_1$ sheets in contrast to the bands forming $\beta_2$ sheets (see Fig.~\ref{fig_EkFS}).

\textbf{4. Superconducting state.}
On the background of the nematic state, we seek for a superconducting state by solving a linearized equation for the order parameter $\Delta_{\vk} = \Delta_0 g_{\vk}$ written as an equation on the eigenvalues $\lambda$ and eigenvectors $g_{\vk}$~\cite{BerkSchrieffer,ScalapinoHF,Graser2009,Korshunov2014eng,KorshunovBook2018},
\begin{equation}
 \lambda g_{\vk} = -\sum\limits_{\nu}\oint\limits_{\nu} \frac{d\vk'_{||}}{2\pi} \frac{1}{2\pi v_{F\vk'}} \tilde\Gamma^{\mu\nu}(\vk,\vk') g_{\vk'},
 \label{eq:Delta}
\end{equation}
where $v_{F\vk'}$ is the Fermi velocity, the contour integral is taken over the parallel to the $\nu$-th Fermi surface sheet component of momentum $\vk'_{||}$, and the band $\mu$ is unambiguously determined by the location of the momentum $\vk$. Positive $\lambda$'s correspond to attraction and the maximal one of them represents the state with the highest critical temperature $T_c$, i.e., the most favorable superconducting state with the corresponding gap function determined by $g_{\vk}$.

Cooper vertex $\tilde\Gamma^{\mu\nu}(\vk,\vk')$ depends on the Coulomb parameters $U$, $U'$, $J$, $J'$ and on $Re\chi(\q,\omega=0)$, see Ref.~\cite{Korshunov2014eng}. Under the spin-rotational invariance assumed here, $U'=U-2J$ and $J'=J$. We vary the other two parameters, $U$ and $J$, of the Hubbard interaction. For the further study we choose the following set of interactions (values in eV)\\
1: $U=1$, $J=0$;
2: $U=1.1$, $J=0$;
3: $U=1$, $J=0.1$;
4: $U=1.2$, $J=0$;
5: $U=1.1$, $J=0.1$;
6: $U=1.2$, $J=0.1$;
7: $U=1$, $J=0.2$;
8: $U=1.3$, $J=0$;
9: $U=1.1$, $J=0.2$;
10: $U=1.3$, $J=0.1$;
11: $U=1.4$, $J=0$;
12: $U=1.2$, $J=0.2$;
13: $U=1.4$, $J=0.1$;
14: $U=1$, $J=0.3$;
15: $U=1.4$, $J=0.15$;
16: $U=1.3$, $J=0.2$.

\begin{figure}
\centering
\includegraphics[width=0.49\linewidth]{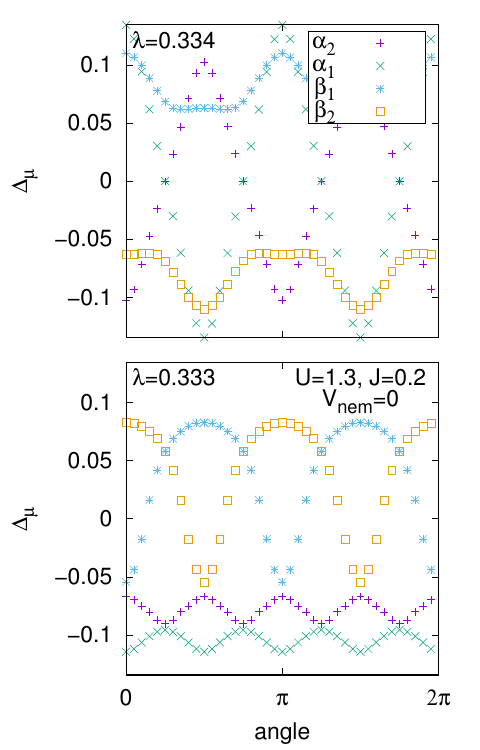}
\includegraphics[width=0.49\linewidth]{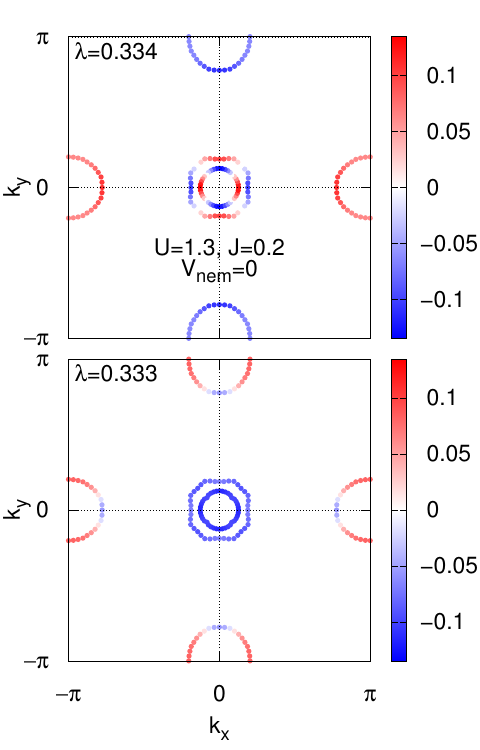}
\caption{Fig.~\ref{fig_gk_Vnem0}. Superconducting order parameter for $U=1.3$, $J=0.2$, and $V_{nem}=0$ for the two leading eigenvalues $\lambda$ at the Fermi surface sheets ($\alpha_{1,2}$, $\beta_{1,2}$). Left: angular dependence at each sheet, right: magnitude of the order parameter is shown as the intensity within the Brillouin zone. All values are in eV. \label{fig_gk_Vnem0}}
\end{figure}

\begin{figure}
\centering
\includegraphics[width=0.49\linewidth]{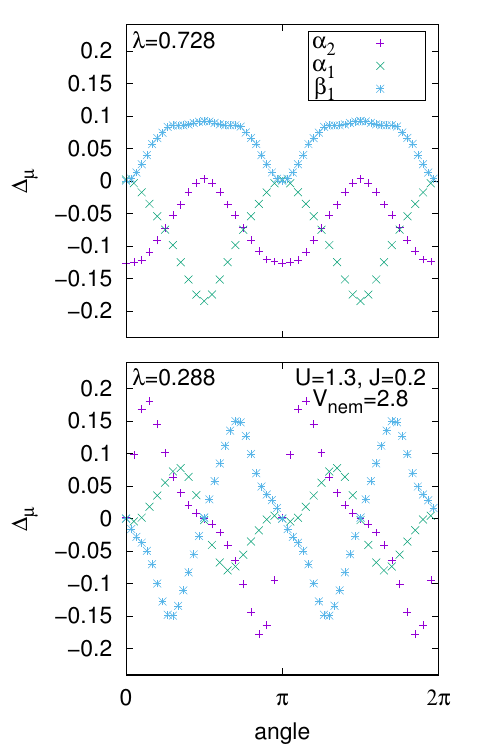}
\includegraphics[width=0.49\linewidth]{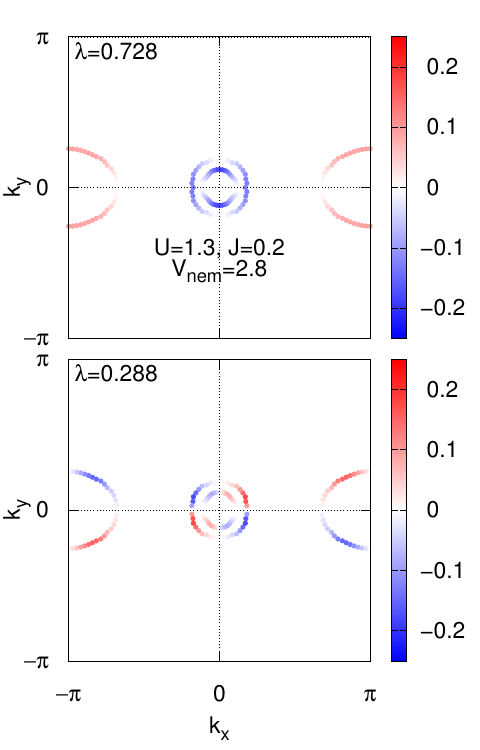}
\caption{Fig.~\ref{fig_gk_Vnem280}. The same as in Fig.~\ref{fig_gk_Vnem0} but for $V_{nem}=2.8$ eV. \label{fig_gk_Vnem280}}
\end{figure}

For the set \#16 ($U=1.3$, $J=0.2$) in Figs.~\ref{fig_gk_Vnem0}-\ref{fig_gk_Vnem280}, we demonstrate the two solutions of Eq.~(\ref{eq:Delta}) with the maximal values of $\lambda$ (leading solutions) in regions I and II: without nematicity at $V_{nem}=0$ in Fig.~\ref{fig_gk_Vnem0} and in the nematic phase at $V_{nem}=2.8$~eV in Fig.~\ref{fig_gk_Vnem280}.

For the current set of interaction parameters, $d_{x^2-y^2}$ and $s_\pm$ type solutions compete in the region I (i.e., they have close values of $\lambda$), however, $d_{x^2-y^2}$ type is winning. In the nematic phase, region II, we can not use the classification of the gaps according to the irreducible representations of the tetragonal phase. Still, we call the state with the larger value of $\lambda$ in Fig.~\ref{fig_gk_Vnem280} as `$s_{\pi\pm}$' to emphasize its connection to the $s_\pm$ state in the tetragonal phase and to point out that the corresponding $\Delta_{\vk}$ is invariant under the rotation by $\pi$, not $\pi/2$ as was for the extended $s$-type symmetry. State with the smaller $\lambda$ in that figure resembles $d_{xy}$ type symmetry, thus we call it like that.

\begin{figure}
\centering
\includegraphics[width=1.0\linewidth]{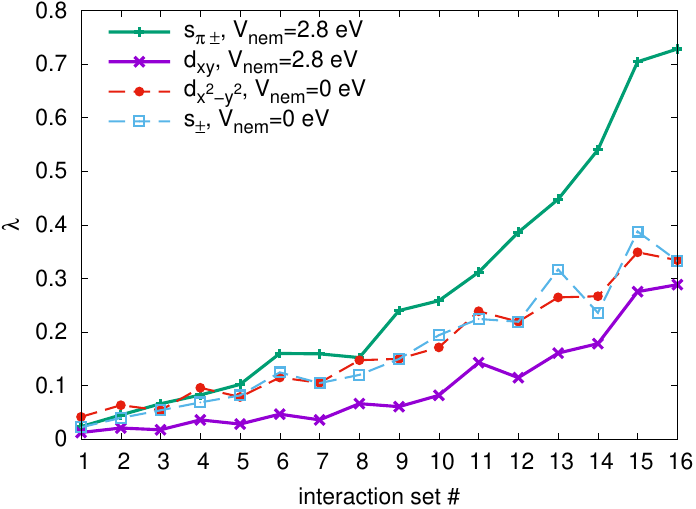}
\caption{Fig.~\ref{fig_lambda}. Leading eigenvalues $\lambda$ for the two values of the nematic interaction coefficient $V_{nem}$ for different sets of Coulomb parameters. \label{fig_lambda}}
\end{figure}

Combined graph of $\lambda$ values for the different sets of onsite Coulomb interactions at $V_{nem}=0$ and at $V_{nem}=2.8$~eV is shown in Fig.~\ref{fig_lambda}. Curves correspond to different calculated gap symmetries. Note that the `nematic type' $s_{\pi\pm}$ always has a larger value of $\lambda$ than the $s_{\pm}$ solution that compete with the $d_{x^2-y^2}$ type. Thus, $T_c$ of the superconducting state coexisting with the nematic state is higher than $T_c$ of the sole superconducting state. This supports the conclusion that the nematic superconducting phase may be more favourable than the state with the unbroken $C_4$ symmetry.

\textbf{5. Conclusion.}
We considered the emergence of the superconductivity on the background of the nematic order in the five-orbital model for iron pnictides and chalcogenides. Nematic order is treated within the mean-field theory with the nematic interaction coefficient $V_{nem}$. Self-consistently calculated nematic order parameter $\Phi_l$ is zero (region I) for values of $V_{nem}$ from zero to some critical value, after which the $\Phi_l$ component corresponding to the orbital $l=d_{xy}$ unevenly becomes finite (region II). Superconducting solution within the spin-fluctuation paring theory is found in both regions, I and II. In the absence of the nematic order (region I), the superconducting order parameter has the $s_\pm$ and $d_{x^2-y^2}$ type structures for the two leading competing solutions. In the nematic phase (region II), two leading solutions are of $s_{\pi\pm}$ and $d_{xy}$ types, moreover, first one always wins. Estimation of the corresponding critical temperatures $T_c$ leads to the conclusion that the nematic superconducting state with the $s_{\pi\pm}$ structure will have the higher $T_c$ than the usual $s_\pm$ and $d_{x^2-y^2}$ states with the unbroken $C_4$ symmetry.

\textbf{Funding.}
This work was supported within the state assignment of Kirensky Institute of Physics.

\textbf{Conflict of interest.}
The authors declare no conflict of interest.




\providecommand{\url}[1]{\texttt{#1}}
\providecommand{\urlprefix}{URL }
\providecommand{\eprint}[2][]{\url{#2}}
\bibliographystyle{jetpl}


\begin{thebibliography}{10}
\providecommand{\url}[1]{\texttt{#1}}
\providecommand{\urlprefix}{URL }
\providecommand{\eprint}[2][]{\url{#2}}

\bibitem{SadovskiiReview2008}
M.~V. Sadovskii,
  \href{https://ufn.ru/en/articles/2008/12/b/}{\href{http://dx.doi.org/10.1070/PU2008v051n12ABEH006820}{Phys.
  Usp.}} \textbf{51}, 1201 (2008).

\bibitem{IzyumovReview2008}
Yu.~A. Izyumov, E.~Z. Kurmaev,
  \href{https://ufn.ru/en/articles/2008/12/d/}{\href{http://dx.doi.org/10.1070/PU2008v051n12ABEH006733}{Phys.
  Usp.}} \textbf{51}, 1261 (2008).

\bibitem{HirschfeldKorshunov2011}
P.~J. Hirschfeld, M.~M. Korshunov, I.~I. Mazin,
  \href{http://stacks.iop.org/0034-4885/74/i=12/a=124508}{Reports on Progress
  in Physics} \textbf{74}, 124508 (2011).

\bibitem{Korshunov2014eng}
M~M Korshunov,
  \href{http://stacks.iop.org/1063-7869/57/i=8/a=813}{\href{http://dx.doi.org/10.3367/UFNe.0184.201408h.0882}{Physics-Uspekhi}}
  \textbf{57}, 813 (2014).

\bibitem{Sadovskii2016}
M.~V. Sadovskii,
  \href{https://ufn.ru/en/articles/2016/10/b/}{\href{http://dx.doi.org/10.3367/UFNe.2016.06.037825}{Phys.
  Usp.}} \textbf{59}, 947 (2016).

\bibitem{MaitiKorshunovPRL2011}
S.~Maiti, M.~M. Korshunov, T.~A. Maier, P.~J. Hirschfeld, A.~V. Chubukov,
  \href{http://link.aps.org/doi/10.1103/PhysRevLett.107.147002}{\href{http://dx.doi.org/10.1103/PhysRevLett.107.147002}{Phys.
  Rev. Lett.}} \textbf{107}, 147002 (2011).

\bibitem{KorshunovEreminResonance2008}
M.~M. Korshunov, I.~Eremin,
  \href{http://link.aps.org/doi/10.1103/PhysRevB.78.140509}{\href{http://dx.doi.org/10.1103/PhysRevB.78.140509}{Phys.
  Rev. B}} \textbf{78}, 140509 (2008).

\bibitem{Maier2008}
T.~A. Maier, D.~J. Scalapino,
  \href{http://link.aps.org/doi/10.1103/PhysRevB.78.020514}{\href{http://dx.doi.org/10.1103/PhysRevB.78.020514}{Phys.
  Rev. B}} \textbf{78}, 020514 (2008).

\bibitem{Korshunov2018}
M.~M. Korshunov,
  \href{https://link.aps.org/doi/10.1103/PhysRevB.98.104510}{\href{http://dx.doi.org/10.1103/PhysRevB.98.104510}{Phys.
  Rev. B}} \textbf{98}, 104510 (2018).

\bibitem{LumsdenReview}
M.~D. Lumsden, A.~D. Christianson,
  \href{http://stacks.iop.org/0953-8984/22/i=20/a=203203}{Journal of Physics:
  Condensed Matter} \textbf{22}, 203203 (2010).

\bibitem{Dai2015}
Pengcheng Dai,
  \href{https://link.aps.org/doi/10.1103/RevModPhys.87.855}{\href{http://dx.doi.org/10.1103/RevModPhys.87.855}{Rev.
  Mod. Phys.}} \textbf{87}, 855 (2015).

\bibitem{Inosov2016}
D.S. Inosov,
  \href{http://www.sciencedirect.com/science/article/pii/S1631070515000523}{\href{http://dx.doi.org/10.1016/j.crhy.2015.03.001}{Comptes
  Rendus Physique}} \textbf{17}, 60  (2016).

\bibitem{KorshunovKuzmichev2022}
Maxim~M. Korshunov, Svetoslav~A. Kuzmichev, Tatiana~E. Kuzmicheva,
  \href{https://www.mdpi.com/1996-1944/15/17/6120}{\href{http://dx.doi.org/10.3390/ma15176120}{Materials}}
  \textbf{15}, 6120 (2022).

\bibitem{Chu2010}
Jiun-Haw Chu, James~G. Analytis, Kristiaan~De Greve, Peter~L. McMahon, Zahirul
  Islam, Yoshihisa Yamamoto, Ian~R. Fisher,
  \href{https://www.science.org/doi/abs/10.1126/science.1190482}{\href{http://dx.doi.org/10.1126/science.1190482}{Science}}
  \textbf{329}, 824 (2010).

\bibitem{Fernandes2012}
R.~M. Fernandes, A.~V. Chubukov, J.~Knolle, I.~Eremin, J.~Schmalian,
  \href{https://link.aps.org/doi/10.1103/PhysRevB.85.024534}{\href{http://dx.doi.org/10.1103/PhysRevB.85.024534}{Phys.
  Rev. B}} \textbf{85}, 024534 (2012).

\bibitem{Fernandes2014}
R.~M. Fernandes, A.~V. Chubukov, J.~Schmalian,
  \href{http://dx.doi.org/10.1038/nphys2877}{Nat. Phys.} \textbf{10}, 97
  (2014).

\bibitem{Li2017}
Jun Li, Paulo~J. Pereira, Jie Yuan et~al.,
  \href{https://doi.org/10.1038/s41467-017-02016-y}{\href{http://dx.doi.org/10.1038/s41467-017-02016-y}{Nature
  Communications}} \textbf{8}, 1880 (2017).

\bibitem{Wang2021}
Xiang Zhou Yifei Li Bolun Teng Peng Dong Jiadian He Yiwen Zhang Yifan~Ding
  Jinghui~Wang, Yueshen~Wu, Jun Li,
  \href{https://doi.org/10.1080/23746149.2021.1878931}{\href{http://dx.doi.org/10.1080/23746149.2021.1878931}{Advances
  in Physics: X}} \textbf{6}, 1878931 (2021).

\bibitem{Sprau2017}
P.~O. Sprau, A.~Kostin, A.~Kreisel, A.~E. Bohmer, V.~Taufour, P.~C. Canfield,
  S.~Mukherjee, P.~J. Hirschfeld, B.~M. Andersen, J.~C.~Seamus Davis,
  \href{https://www.science.org/doi/abs/10.1126/science.aal1575}{\href{http://dx.doi.org/10.1126/science.aal1575}{Science}}
  \textbf{357}, 75 (2017).

\bibitem{Kuroki2008}
Kazuhiko Kuroki, Seiichiro Onari, Ryotaro Arita, Hidetomo Usui, Yukio Tanaka,
  Hiroshi Kontani, Hideo Aoki,
  \href{http://link.aps.org/doi/10.1103/PhysRevLett.101.087004}{\href{http://dx.doi.org/10.1103/PhysRevLett.101.087004}{Phys.
  Rev. Lett.}} \textbf{101}, 087004 (2008).

\bibitem{Graser2009}
S.~Graser, T.A. Maier, P.J. Hirschfeld, D.J. Scalapino,
  \href{http://stacks.iop.org/1367-2630/11/i=2/a=025016}{New Journal of
  Physics} \textbf{11}, 025016 (2009).

\bibitem{Yamase2005}
Hiroyuki Yamase, Vadim Oganesyan, Walter Metzner,
  \href{https://link.aps.org/doi/10.1103/PhysRevB.72.035114}{\href{http://dx.doi.org/10.1103/PhysRevB.72.035114}{Phys.
  Rev. B}} \textbf{72}, 035114 (2005).

\bibitem{Choudhury2022}
Sourav Sen~Choudhury, Sean Peterson, Yves Idzerda,
  \href{https://link.aps.org/doi/10.1103/PhysRevB.105.214515}{\href{http://dx.doi.org/10.1103/PhysRevB.105.214515}{Phys.
  Rev. B}} \textbf{105}, 214515 (2022).

\bibitem{Castallani1978}
C.~Castellani, C.~R. Natoli, J.~Ranninger,
  \href{http://link.aps.org/doi/10.1103/PhysRevB.18.4945}{\href{http://dx.doi.org/10.1103/PhysRevB.18.4945}{Phys.
  Rev. B}} \textbf{18}, 4945 (1978).

\bibitem{Oles1983}
A.~M. Ole\'{s},
  \href{http://link.aps.org/doi/10.1103/PhysRevB.28.327}{\href{http://dx.doi.org/10.1103/PhysRevB.28.327}{Phys.
  Rev. B}} \textbf{28}, 327 (1983).

\bibitem{BerkSchrieffer}
N.~F. Berk, J.~R. Schrieffer,
  \href{http://link.aps.org/doi/10.1103/PhysRevLett.17.433}{\href{http://dx.doi.org/10.1103/PhysRevLett.17.433}{Phys.
  Rev. Lett.}} \textbf{17}, 433 (1966).

\bibitem{ScalapinoHF}
D.~J. Scalapino, E.~Loh, J.~E. Hirsch,
  \href{http://link.aps.org/doi/10.1103/PhysRevB.34.8190}{\href{http://dx.doi.org/10.1103/PhysRevB.34.8190}{Phys.
  Rev. B}} \textbf{34}, 8190 (1986).

\bibitem{KorshunovBook2018}
M.M. Korshunov, \emph{Perturbation Theory: Advances in Research and
  Applications}  (Nova Science Publishers Inc., New York 2018), chapter
  Itinerant Spin Fluctuations in Iron-Based Superconductors, pp. 61--138.

\end{thebibliography}


\vfill\eject

\end{document}